\documentclass[prb,showpacs,floatfix,twocolumn,superscriptaddress]{revtex4}
\usepackage{graphicx}
\usepackage{dcolumn}
\begin{document}


\title{Optical study of the charge-density-wave mechanism in 2$H$-TaS$_2$ and Na$_x$TaS$_2$}

\author{W. Z. Hu}
\author{G. Li}
\author{J. Yan}
\author{H. H. Wen}
\affiliation{Beijing National Laboratory for Condensed Matter
Physics, Institute of Physics, Chinese Academy of Sciences,
Beijing 100080, P. R. China}
\author{G. Wu}
\author{X. H. Chen}
\affiliation{Hefei  National Laboratory for Physical  Science at
Microscale  and  Department of  Physics, University  of Science
and Technology  of  China, Hefei 230026, P. R. China}

\author{N. L. Wang}
\email{nlwang@aphy.iphy.ac.cn}%
\affiliation{Beijing National Laboratory for Condensed Matter
Physics, Institute of Physics, Chinese Academy of Sciences,
Beijing 100080, P. R. China}
%


\begin{abstract}
We report an optical study of transition metal dichalcogenide
2$H$-TaS$_2$ and Na intercalated superconductor Na$_x$TaS$_2$ over
a broad frequency range at various temperatures. A clear gap
feature was observed for 2$H$-TaS$_2$ when it undergoes a CDW
transition. The existence of a Drude component in
$\sigma$$_1$($\omega$) below T$_{CDW}$ indicates that the Fermi
surface of 2$H$-TaS$_2$ is only partially gapped in CDW state. The
spectral evolution of two different Na$_x$TaS$_2$ crystals further
confirms that the partial gap structure observed in 2$H$-TaS$_2$
has a CDW origin. The CDW mechanism for 2$H$-TaS$_2$, the
competition between CDW and superconductivity in Na$_x$TaS$_2$
system are discussed.
\end{abstract}

\pacs{78.20.-e, 71.45.Lr}

\maketitle

Charge density wave (CDW) order and its coexistence with
superconductivity in 2$H$ type transition metal dichalcogenides
(TMDC) has been one of the major subjects in condensed matter
physics since 1970's. The name "2$H$" is derived from structural
character where "2" represents two chalcogen-metal-chalcogen
sandwiches in one unit cell and "$H$" stands for "hexagonal". In
this family, 2$H$-TaSe$_2$ has an incommensurate followed by a
commensurate CDW transitions at 122K and 90K; while 2$H$-TaS$_2$
and 2$H$-NbSe$_2$ only undergo one incommensurate CDW transition
at 75K and 35K, respectively\cite{Wilson and Di Salvo}. On the
contrary, the superconducting transition temperature T$_c$
increases from 2$H$-TaSe$_2$ over 2$H$-TaS$_2$ to 2$H$-NbSe$_2$,
suggesting that the CDW order and superconductivity compete.
Additionally, upon intercalation of Na into 2$H$-TaS$_2$,
T$_{CDW}$ is suppressed but T$_c$ increases from 0.8 to 4.4 K,
yielding further evidence for the competition between these two
orders\cite{NaxTaS2}. Although it is generally believed that the
superconductivity in 2$H$-TMDC is caused by conventional
electron-phonon coupling, the driving mechanism for CDW transition
is still not well understood. Different viewpoints including Fermi
surface (FS) nesting\cite{Wilson}, saddle point
mechanism\cite{Rice} or other hidden order in \textit{\textbf{k}}
space\cite{Seifarth, Shen} have been proposed to explain the CDW
instability, but no consensus has been reached.

Optical spectroscopy is a powerful bulk sensitive technique with
high energy resolution, but previous researches on the in-plane
optical properties of 2$H$-TaSe$_2$ and 2$H$-NbSe$_2$ did not find
any sharp changes directly related to CDW transition\cite {FT-IR}.
As 2$H$-TaS$_2$ has only one CDW transition with relatively higher
T$_{CDW}$, it is therefore an ideal system to study CDW mechanism
in $2H$-TMDC family. However, no detailed optical research was
performed on 2$H$-TaS$_2$ except some early work on the high
energy structure at room or liquid N$_2$ temperatures\cite
{optic}. In this work, we report the first detailed optical study
of 2$H$-TaS$_2$ and Na$_x$TaS$_2$ over a broad frequency range at
various temperatures to reveal the CDW mechanism. The overall
frequency dependent reflectivity R($\omega$) of 2$H$-TaS$_2$ shows
a metallic response, but a mid-infrared suppression feature
emerges below T$_{CDW}$. The real part of conductivity
$\sigma$$_1$($\omega$) is composed of a Drude response together
with a weak mid-infrared peak in the CDW state, indicating a
partial gap was formed on the Fermi surface. For Na$_x$TaS$_2$,
CDW features turn weaker and tend to disappear as Na content
increases. All those optical results, being consistent with the
transport measurements obtained on the same crystals, can be well
understood in the frame of FS nesting mechanism.

\begin{figure}[b]
\includegraphics[width=6cm]{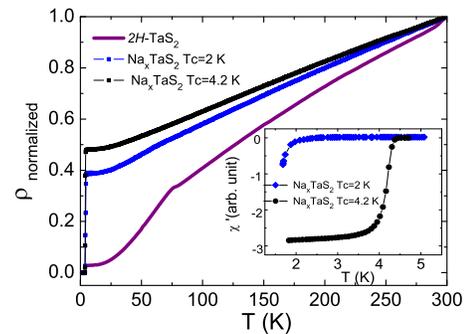}%
\vspace*{-0.50cm}%
\caption{\label{fig:resistivity}(color online) T-dependent
resistivity for 2$H$-TaS$_2$ and two Na$_x$TaS$_2$ crystals (all
normalized to values at 300 K). The inset shows the real part of
ac susceptibility below 5K, which identifies T$_c$ for two
Na$_x$TaS$_2$ crystals 2 K and 4.2 K, respectively.}
\end{figure}

2$H$-TaS$_2$ single crystals were grown by chemical iodine-vapor
transport method. High purity Ta (99.95\%) and S (99.5\%) powder
were mixed, thoroughly ground, pressed into a pellet, and sealed
under vacuum in a quartz tube (\O 13 mm$\times$150 mm) with iodine
(10 mg/cm$^{-3}$). Single crystals were obtained after growing at
720-700$^{\circ}$C for a week and cooling down slowly to room
temperature. 2$H$ type Na$_x$TaS$_2$ single crystals were grown by
chemical reaction, as described previously\cite{NaxTaS2}. The
T-dependent resistivity was obtained by four contacts technique in
a Quantum Design PPMS and the results are shown in Fig. 1. There
is a clear CDW transition at 75K for 2$H$-TaS$_2$, while such
feature is vague around 65K for Na$_x$TaS$_2$ T$_c$=2 K sample,
and totally disappears in the higher Na content crystals
(T$_c$=4.2 K). The superconducting transition temperature for
Na$_x$TaS$_2$ was identified by ac susceptibility as shown in the
inset.

Optical experiments were carried out on as-grown surfaces by a
Bruker IFS 66v/s spectrometer in the range from 30 to 25,000
cm$^{-1}$ with \emph{in-situ} overcoating technique to get the
in-plane reflectivity. Considering the small size of our samples
(about $1.5\times2$ mm$^{2}$), data below 150 cm$^{-1}$ are cut
off for reliability. Fig. 2 shows the T-dependent R($\omega$) for
2$H$-TaS$_2$ and Na$_x$TaS$_2$, both display a metallic behavior,
including the high reflectance at low $\omega$, a sharp drop in
the near-infrared region (plasma edge), and interband transitions
at higher frequencies. The difference between the pure and
intercalated materials in low energies is evident. For
2$H$-TaS$_2$, R($\omega$) at 10 K and 60 K are suppressed below
the curve of 90 K from 400 to 5000 cm$^{-1}$. This is a typical
partial-gap behavior on Fermi surface manifested in
reflectance\cite{Wang}. Similar suppression is weak in
Na$_x$TaS$_2$ T$_c$=2 K sample, and almost invisible in T$_c$=4.2
K sample with higher Na content.

\begin{figure}[t]
\includegraphics[width=2.7in]{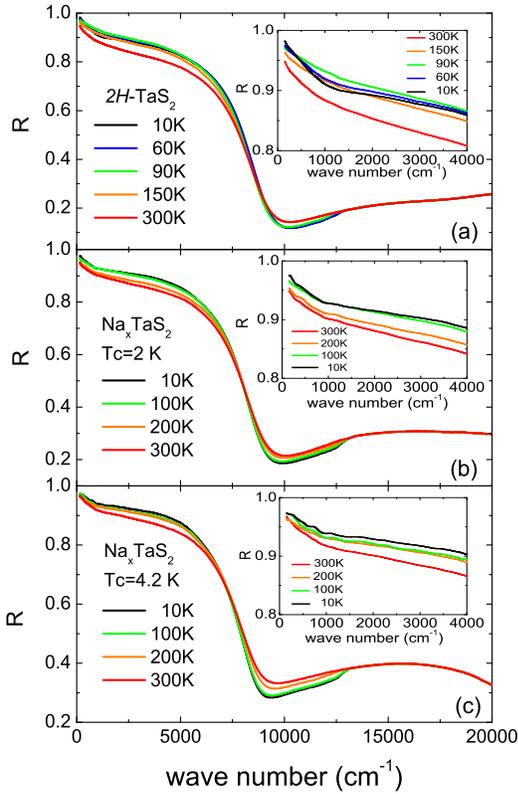}%
\vspace*{-0.20cm}%
\caption{\label{fig:ref cmp}(color online) Reflectivity of (a)
2$H$-TaS$_2$ (b) Na$_x$TaS$_2$ (T$_c$=2 K) (c)Na$_x$TaS$_2$
(T$_c$=4.2 K) from 150 cm$^{-1}$ to 20000 cm$^{-1}$ at various
temperatures. The inset figures focus on the spectra from 150
cm$^{-1}$ to 4000 cm$^{-1}$.}
\end{figure}

Fig. 3 illustrates the real part of conductivity obtained by
Kramers-Kronig transformation of R($\omega$) at selected
temperatures. The inset shows $\sigma$$_1$($\omega$) below 4000
cm$^{-1}$ at 10K. The existence of the Drude component at all
temperatures in both 2$H$-TaS$_2$ and Na$_x$TaS$_2$ confirms their
metallic behavior. For 2$H$-TaS$_2$, a mid-infrared peak emerges
below T$_{CDW}$ and vanishes at higher T. In order to clarify this
feature, we use Drude model to fit the low frequency part of 10K's
curve and subtract this free-carrier contribution from the
experiment result, as shown in inset of Fig. 3 (a). Then we get a
gap of 45 ${meV}$, close to reported CDW-gap value of 50 ${meV}$
by STM experiment\cite{STM}. As the scattering rate 1/$\tau$ in
simple Drude model is a frequency independent constant, such
fitting is just a qualitative analysis. In Na$_x$TaS$_2$ case,
similar gap-like feature is obscure and shifts to lower energy for
Na$_x$TaS$_2$ T$_c$=2 K sample, and almost disappears in the
higher Na content crystal. Since the mid-infrared peak in
$\sigma$$_1$($\omega$) appears only below T$_{CDW}$ for
CDW-bearing samples, and vanishes in Na$_x$TaS$_2$ with no CDW
transition, which consist with dc transport results, thus this
peak in $\sigma$$_1$($\omega$) has a CDW origin. As the Drude
response coexists with the mid-infrared peak below T$_{CDW}$, the
Fermi surface is only partially gapped in the CDW state.

\begin{figure}[t]
\includegraphics[width=2.7in]{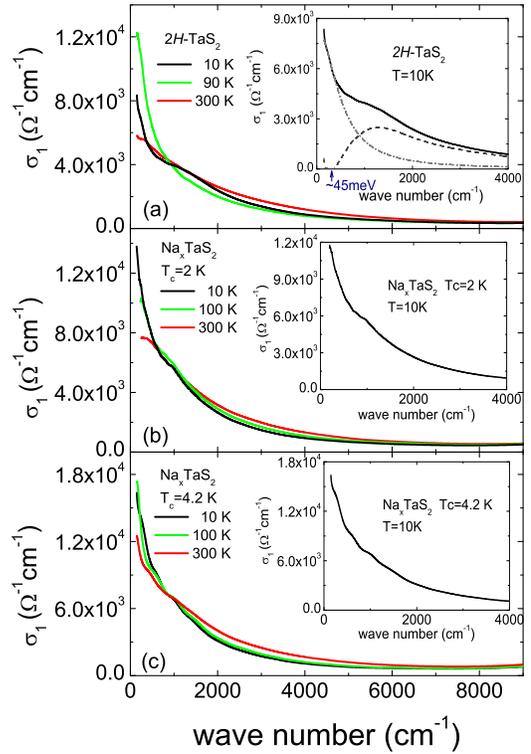}%
\vspace*{-0.20cm}%
\caption{\label{fig:S1 cmp}(color online) The real part of
conductivity of (a)2$H$-TaS$_2$ (b)Na$_x$TaS$_2$ (T$_c$=2 K) (c)
Na$_x$TaS$_2$ (T$_c$=4.2 K) from 150 cm$^{-1}$ to 9000 cm$^{-1}$
at selected temperatures. The inset figures amplified the spectra
at 10K from 150 cm$^{-1}$ to 4000 cm$^{-1}$. For 2$H$-TaS$_2$, the
experiment result, a Drude fit, and $\sigma$$_1$($\omega$)
subtracts Drude fit is plotted by black line, gray dash dot line,
dark gray dash line, respectively.}
\end{figure}

It is well known that a partial gap structure can be well resolved
in the spectrum of the frequency dependent scattering rate
$\Gamma(\omega)$\cite{Wang,Basov}. Fig. 4 shows the
$\Gamma(\omega)$ spectra for those samples obtained from the
extended Drude model $\Gamma(\omega)$=$(\omega_p^2 /
4\pi)$Re[1/$(\sigma(\omega)]$, where $\omega_p$ is the overall
plasma frequency and can be obtained by summarizing
$\sigma_1(\omega)$ up to the reflectance edge frequency. The
obtained $\omega_p$ values are roughly 2.4$\times$10$^4$ cm$^{-1}$
for 2$H$-TaS$_2$, 2.6$\times$10$^4$ cm$^{-1}$ for Na$_x$TaS$_2$
T$_c$=2 K sample and 3.0$\times$10$^4$ cm$^{-1}$ for Na$_x$TaS$_2$
T$_c$=4.2 K sample, respectively. For 2$H$-TaS$_2$, above
T$_{CDW}$, $\Gamma(\omega$) is almost unchanged in shape but
decreases in magnitude as T decreases. In the CDW state, a broad
peak forms around 1000 cm$^{-1}$ and gets stronger at 10K. For
Na$_x$TaS$_2$, similar feature is weak and shifts slightly to
lower energy, and tends to vanish eventually at higher Na content.
We note that the spectral feature in $\Gamma(\omega$) is very
similar to the behavior in other materials with a partially gaped
Fermi surface, such as the antiferromagnet Cr where a
spin-density-wave gap opens on parts of the FS \cite{Basov}, and
the electron-doped high-T$_c$ cuprate Nd$_{2-x}$Ce$_x$CuO$_4$ in
which a spin-correlation gap appears at "hot spots" (the
intersecting points of the FS with the antiferromagnetic zone
boundary)\cite{Wang}. In comparison with those work, the
$\Gamma(\omega$) spectra shown in Fig. 4 further clarifies the
existence of CDW partial gap on the FS in 2$H$-TaS$_2$ and its
vanishing tendency upon Na-intercalation. We emphasize that the
CDW gap feature in optics is weak in 2$H$-TaS$_2$. Early work
based on other technique indicated that only 10 to 20$\%$ FS were
removed by CDW gap in $2H$-Ta materials\cite{Wilson}.
Qualitatively, our result is consistent with those work.

\begin{figure}[t]
\includegraphics[width=2.2in]{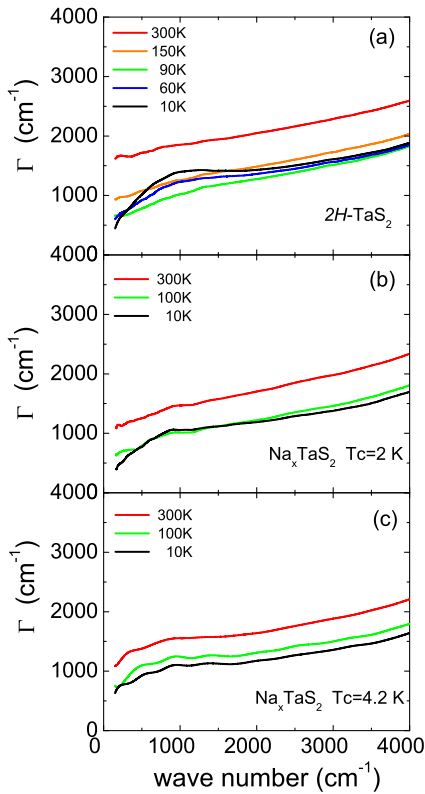}%
\vspace*{-0.20cm}%
\caption{\label{fig:G1 cmp}(color online) The scattering rate of
(a)2$H$-TaS$_2$ (b)Na$_x$TaS$_2$ (T$_c$=2 K) (c)Na$_x$TaS$_2$
(T$_c$=4.2 K)}
\end{figure}

As shown above, we identified the formation of CDW induced partial
gap and its evolution with Na intercalation. Before giving further
discussion on charge-density-wave mechanism, a brief review about
ARPES results is necessary. Previous studies on
2$H$-NbSe$_2$\cite{Kiss, Straub NbSe2, Tonjes, Rossnagel NbSe2},
2$H$-TaS$_2$\cite{Tonjes} and 2$H$-TaSe$_2$\cite{Liu TaSe2 00,
Rossnagel TaSe2} showed that they all have similar Fermi surface,
i.e., two double-wall hole pockets around $\Gamma$ and $K$.
Especially, a recent research indicates that Na$_x$TaS$_2$ also
has such two-pockets Fermi surface\cite{Shen}. Apart from
observations of FS topology, the existence of saddle point at 1/2
$\Gamma$$K$\cite{Straub NbSe2}, or extended saddle band along
$\Gamma$$K$\cite{Liu TaSe2 98, Liu TaSe2 00, Tonjes} were
confirmed by ARPES for 2$H$-TMDC.

The key requirement for any CDW mechanism is the direct
observation of a CDW gap, and a match between the nesting vector
which links the gapped region and the CDW vector associated with
the lattice superstructure. Fermi surface nesting is a traditional
mechanism for CDW transitions, which predicts the opening of
charge-density-wave gap at parallel regions of Fermi surface
connected by CDW vector. For 2$H$-TMDC members, the possibility
about $\Gamma$ pocket self nesting at parallel hexagon
boundaries\cite{Straub NbSe2} has been excluded because the
distance between two nesting regions does not match the length of
CDW vector associated with the nearly $3\times3$ superlattice, and
no CDW gap was found around $\Gamma$ pocket\cite{Liu TaSe2 00, Liu
TaSe2 98, Tonjes, Rossnagel NbSe2, Rossnagel TaSe2, Valla TaSe2,
Shen}. On the other hand, there are evidences for a weak nesting
at Fermi surface around $K$\cite{Wilson, Wexler, Rossnagel NbSe2,
Rossnagel TaSe2}, and the energy gap around inner $K$ pocket below
T$_{CDW}$ was clearly observed in 2$H$-TaSe$_2$\cite{Liu TaSe2 00,
Rossnagel TaSe2} and 2$H$-TaS$_2$\cite{Tonjes}. Besides FS nesting
scenario, saddle point mechanism is another traditional model
which regards the saddle points close to E$_F$ as scattering
"sink" and their removal in the CDW state will enhance dc
conductivity\cite{Rice}. But saddle point's position observed by
ARPES would lead to a $2\times2$ superlattice, inconsistent with
the nearly $3\times3$ superlattice for 2$H$-TMDC in the CDW state.
Moreover, the saddle band is not so close to the Fermi
energy\cite{Straub NbSe2, Tonjes, Rossnagel NbSe2} as Rice and
Scott predicted. Therefore, some recent ARPES studies try to find
other "hidden" order in \textit{\textbf{k}} space to explain the
CDW instability, such as the CDW-vector-matched saddle points
where the "gap" follows BCS description in CDW
state\cite{Seifarth} or special "gapped" regions around $M$ point
of the Brillouin zone\cite{Shen}. In fact, $M$ is close to $K$
pocket, and "gap" appears around $M$ is usually accompanied by the
opening of a real CDW gap on FS around $K$ pocket. Specially, for
energy band which does not across E$_F$, the observation of band
dispersion anomaly is not a real "gap" but an "energy
shift"\cite{Liu TaSe2 98, Liu TaSe2 00}, so "gapped" region around
$M$ point cannot account for charge-density-wave instability in
2$H$-TMDC.

Optical probe offers supplementary information on charge
excitations with high energy resolution, and its result should be
consistent with ARPES observation. On this basis, our data can be
naturally understood in the frame of Fermi surface nesting
mechanism. First of all, we observe the temperature evolution of a
partial gap formed on Fermi surface, which is directly related to
the onset of charge-density-wave transition. For 2$H$-TaS$_2$,
both the suppression in R($\omega$), the mid infrared peak in
$\sigma$$_1$($\omega$), and the broad peak in $\Gamma$($\omega$)
below T$_{CDW}$ testify the existence of a gapped Fermi surface in
CDW state. In addition, all those special characters turn weaker
after Na intercalation, and tend to disappear with increasing Na
content, which again convince that these characters have a CDW
origin. As the Drude component in $\sigma$$_1$($\omega$) never
disappears at all temperatures, the free carrier contribution
exists in the CDW state, consistent with the fact that $\Gamma$
pocket is unaffected in the CDW state for both
2$H$-TaS$_2$\cite{Tonjes} and low Na content
Na$_x$TaS$_2$\cite{Shen}. ARPES experiments indicate that the
$\Gamma$ pocket dominates the transport properties\cite{Valla
TaSe2}, formation of a CDW gap at $K$ pocket, which plays a less
role in transport above T$_{CDW}$, would reduce the scattering
channels, therefore enhance the metallic behavior in CDW state,
thus the resistivity drops after CDW transition for 2$H$-TaS$_2$
and Na$_x$TaS$_2$ T$_c$=2 K samples. The FS nesting mechanism can
also help in understanding why Na intercalation suppresses CDW
transition and increases T$_c$. As evidenced by ARPES experiments,
both the inner and outer $K$ pockets contribute to
superconductivity by opening up superconducting gaps below T$_c$
in those region which were not gapped by CDW
ordering\cite{Yokoya,Valla NbSe2}. Here we should keep in mind
that only certain regions of inner $K$ pocket satisfy the CDW
nesting condition\cite{Wilson, Wexler, Rossnagel NbSe2, Rossnagel
TaSe2}. With Na intercalation, the shapes of Fermi surfaces
surrounding both $\Gamma$ and $K$ are expected to change. This was
confirmed by recent ARPES research on Na$_x$TaS$_2$
crystals\cite{Shen}. Then it is possible that some regions of $K$
pocket would no longer be nested upon Na intercalations, leading
to a gradual reduction of gapped region at Fermi surface,
therefore the CDW gap feature in $\sigma$$_1$($\omega$) and
$\Gamma$($\omega$) for Na$_x$TaS$_2$ crystals are weaker than that
of 2$H$-TaS$_2$. As fewer region at the K pocket can be gapped in
the CDW state, density of states at Fermi energy would be larger
than the case in 2$H$-TaS$_2$, following BCS theory, T$_c$ will
increase as a result.

In conclusion, we report the optical study of 2$H$-TaS$_2$ and
Na$_x$TaS$_2$ over a broad frequency range at various
temperatures. Both the mid-infrared suppression in R($\omega$),
the peak in $\sigma$$_1$($\omega$) and $\Gamma$($\omega$) appear
in accord with the onset of CDW transition, signaling the
formation of a partial gap on Fermi surface in the CDW state. The
gradual removal of CDW character for two different Na$_x$TaS$_2$
samples further confirm the gap-like features observed in
2$H$-TaS$_2$ have a CDW origin. The charge-density-wave mechanism
in 2$H$-TMDC and Na's role in the suppression of CDW can be
understood in the frame of Fermi surface nesting mechanism.
Specifically, we suggest the $K$ pocket is not only responsible
for CDW instability, but also contributes largely to
superconductivity.

This work is supported by National Science Foundation of China,
the Knowledge Innovation Project of Chinese Academy of Sciences,
and the 973 project of Ministry of Science and Technology of
China.


\begin{references}

\bibitem{Wilson and Di Salvo} J. A. Wilson and A. D. Yoffe, Adv. Phys. \textbf{18}, 193 (1969); J. A. Wilson, F. J. Di Salvo and S. Mahajan, Adv. Phys.\textbf{24}, 117 (1975); D. E. Moncton, J. D. Axe, and F. J. DiSalvo, Phys. Rev. B
\textbf{16}, 801 (1977).

\bibitem{NaxTaS2} L. Fang, Y. Wang, P. Y. Zou, L. Tang, Z. Xu, H. Chen,
C. Dong, L. Shan, and H. H. Wen, Phys. Rev. B \textbf{72}, 014534
(2005).

\bibitem{Wilson} J. A. Wilson, Phys. Rev. B \textbf{15}, 5748 (1977).

\bibitem{Rice} T. M. Rice and G. K. Scott, Phys. Rev. lett. \textbf{35}, 120 (1975).

\bibitem{Seifarth} O. Seifarth, S. Gliemann, M. Skibowski, L. Kipp, Journal of Electron Spectroscopy and Related Phenomena \textbf{137-140}, 675 (2004).

\bibitem{Shen} D. W. Shen, B. P. Xie, J. F. Zhao, L. X. Yang, L. Fang, J. Shi, R. H. He, D. H. Lu, H. H. Wen, and D. L. Feng,
cond-mat/0612064.

\bibitem{FT-IR} A. S. Barker, Jr., J. A. Ditzenberger, and F. J. Di Salvo, Phys. Rev. B \textbf{12}, 2049 (1975); V. Vescoli, L. Degiorgi H. Berger and L. Forr¨®, Phys. Rev. lett. \textbf{81}, 453
(1998); S. V. Dordevic, D. N. Basov, R. C. Dynes, and E. Bucher,
Phys. Rev. B \textbf{64}, 161103(R) (2001); S. V. Dordevic, D. N.
Basov, R. C. Dynes, B. Ruzicka, V. Vescoli, L. Degiorgi, H.
Berger, R. Ga¨¢l, L. Forr¨®, and E. Bucher, Eur. Phys. J. B
\textbf{33}, 15 (2003).

\bibitem{optic} A. R. Beal, H. P. Hughes and W. Y. Liang J. Phys. C: Solid State Phys. \textbf{8}, 4236
(1975); S. S. P. Pakkin and A. R. Beal, Philo. Mag. B \textbf{42},
627 (1980).

\bibitem{Wang} N. L. Wang, G. Li, Dong Wu, X. H. Chen, C. H. Wang, and H. Ding, Phys. Rev. B \textbf{73}, 184502 (2006).

\bibitem{STM} C. Wang, B. Giambattista, C. G. Slough, and
R. V. Coleman, Phys. Rev. B \textbf{42}, 8890 (1990).

\bibitem{Basov} D. N. Basov, E. J. Singley, and S. V. Dordevic, Phys. Rev. B \textbf{65}, 054516 (2002).

\bibitem{Kiss} T. Kiss, T. Yokoya, A. Chainani, S. Shin, M. Nohara, H. Takagi, Physica B \textbf{312-313}, 666 (2002)

\bibitem{Straub NbSe2} Th. Straub, Th. Finteis, R. Claessen, P. Steiner, S. H¨¹fner, P. Blaha, C. S. Oglesby, and E. Bucher, Phys. Rev. lett. \textbf{82}, 4504 (1999).

\bibitem{Tonjes} W. C. Tonjes, V. A. Greanya, R. Liu, C. G. Olson, and P. Molinie, Phys. Rev. B \textbf{63}, 235101 (2001).

\bibitem{Rossnagel NbSe2} K. Rossnagel, O. Seifarth, L. Kipp, M. Skibowski, D. Vo\ss, P. Kruger, A. Mazur, and
J. Pollmann, Phys. Rev. B \textbf{64}, 235119 (2001).

\bibitem{Liu TaSe2 00} R. Liu, W. C. Tonjes, V. A. Greanya, C. G. Olson, and R. F. Frindt, Phys. Rev. B \textbf{61}, 52125762 (2000).

\bibitem{Rossnagel TaSe2} K. Rossnagel, Eli Rotenberg, H. Koh, N. V. Smith, and L. Kipp, Phys. Rev. B \textbf{72}, 121103(R) (2005).

\bibitem{Liu TaSe2 98} R. Liu, C. G. Olson, W. C. Tonjes, and R. F. Frindt,  Phys. Rev. lett. \textbf{80}, 5762
(1998).

\bibitem{Valla TaSe2} T. Valla, A.V. Fedorov, P. D. Johnson, J. Xue, K. E. Smith, and F. J.
DiSalvo, Phys. Rev. lett. \textbf{85}, 4759 (2000).

\bibitem{Wexler} G. Wexler and A. M. Woolley, J. Phys. C: Solid State Phys. \textbf{9}, 1185 (1976).

\bibitem{Valla NbSe2} T. Valla, A.V. Fedorov, P. D. Johnson, P-A. Glans, C. McGuinness, K. E. Smith, E.Y. Andrei, H. Berger, Phys. Rev. Lett. \textbf{92},
086401 (2004).

\bibitem{Yokoya} T. Yokoya, T. Kiss, A. Chainani, S. Shin, M. Nohara, H. Takagi, Science \textbf{294}, 2518 (2001).

\end{references}
\end{document}